\documentclass[prl,superscriptaddress,showpacs,amsmath,amssymb,twocolumn,floatfix]{revtex4}
\usepackage{graphicx}% Include figure files
\usepackage{dcolumn}% Align table columns on decimal point
\usepackage{color}
\newcommand{\vc}[1]{{\mathbf #1}}

\begin{document}

\title{ 
Electronic Origin of the Volume Collapse in Cerium
}

\begin{abstract}
The cerium
$\alpha$-$\gamma$ phase transition is characterized by means of
a many-body Jastrow-correlated wave function,
which minimizes the variational energy of the first-principles
scalar-relativistic Hamiltonian, and includes correlation effects in a
non-perturbative way. Our variational ansatz
accurately reproduces the structural properties of the two phases, and
proves that even at temperature $T=0$K the system undergoes a first
order transition, with 
ab-initio parameters which are 
seamlessly connected to the ones 
measured by experiment at finite $T$. We show that the transition is
related to 
a complex rearrangement of the electronic structure,
with key role played by the $p$-$f$ hybridization. The underlying
mechanism unveiled by this work can hold in many Ce-bearing compounds,
and more generally in other f-electron systems.
\end{abstract}

\author{N. Devaux}
\affiliation{Institut de Min\'eralogie, 
de Physique des Mat\'eriaux et de Cosmochimie,
Universit\'e Pierre et Marie Curie,
case 115, 4 place Jussieu, 75252, Paris cedex 05, France}
\author{M. Casula}
\email[]{michele.casula@impmc.upmc.fr}
\affiliation{CNRS and Institut de Min\'eralogie,
de Physique des Mat\'eriaux et de Cosmochimie,
Universit\'e Pierre et Marie Curie,
case 115, 4 place Jussieu, 75252, Paris cedex 05, France}
\author{F. Decremps}
\affiliation{Institut de Min\'eralogie,
de Physique des Mat\'eriaux et de Cosmochimie,
Universit\'e Pierre et Marie Curie,
case 115, 4 place Jussieu, 75252, Paris cedex 05, France}
\author{S. Sorella}
\email[]{sorella@sissa.it}
\affiliation{International School for Advanced Studies (SISSA) Via Beirut 2,4
  34014 Trieste, Italy and INFM Democritos National Simulation Center,
  Trieste, Italy} 
\date{\today}

\pacs{71.15.-m, 71.20.Eh, 71.27.+a, 02.70.Ss}

\maketitle

Understanding the anomalous behavior of cerium, the prototypical
$f$-electron system, is one of the main challenges in condensed matter
physics. The $4f$ electrons are strongly localized and their on-site 
Coulomb repulsion is large compared to bandwidth. Among all
lanthanides, cerium is particularly fascinating, due to
the strong hybridization with the
$6s$−$6p$−$5d$ bands, all present at the Fermi level.
The origin of the cerium volume collapse along the 
isostructural $\alpha$-$\gamma$ transition has been a puzzle 
since its discovery in 1927\cite{bridgman1927}.
A microscopic comprehensive description of the transition is still lacking, because a
direct comparison with the measured structural properties requires an accuracy
below 10 meV. This challenges any ab-initio method, particularly in a
regime of strong correlation. 
Model calculations have been performed in the
Mott\cite{PhysRevLett.74.2335}, Kondo\cite{PhysRevLett.49.1106,allen1992} and 
dynamical mean field theory (DMFT)
\cite{PhysRevLett.87.276403,PhysRevLett.87.276404,PhysRevLett.94.036401,PhysRevLett.96.066402,Amadon2014} 
frameworks,
with input parameters either
chosen ad-hoc or derived from first-principles density functional
theory (DFT) and cRPA calculations\cite{ferdiU}.
Fully first-principle electronic structure schemes, such as
DFT\cite{PhysRevLett.109.146402} or GW\cite{sakuma}, grasp some
features of the $\alpha$ and $\gamma$ phases, but the quantitative
agreement with experiment is generally quite poor.

Experimentally, pure cerium undergoes the 
$\alpha$-$\gamma$ transition 
always at finite temperature $T$. 
Recently, very accurate X-ray diffraction measurements
undoubtedly confirmed the first-order Fm$\bar{3}$m isostructural
character of the 
transition\cite{decremps}. 
The first-order line extrapolates to
zero-$T$ at negative pressures. 
Nevertheless, the
$T$=0K determination of its phase diagram is extremely
important as it can shed light on the underlying electronic structure
mechanism of the 
transition, and clarify some critical points still under debate.
For instance, 
some experiments with cerium alloys seem
to find a critical low-$T$ end-point on the $\alpha$-$\gamma$ phase
boundary\cite{thompson1983}, where the effect of alloying is expected to provide
a negative chemical pressure on the cerium sites. However, it has
also been proven that the end-point of the critical line can be
tuned down to zero $T$ by changing the bulk modulus
through alloying, thus opening the way of new low-T scenarios, like
superconducting and non-Fermi liquid fluctuations\cite{dzero2006}.
The presence of a low-T end-point is obviously material
dependent and it is therefore possible that cerium
 allows instead a genuine 
 $f$-electron driven 0K quantum phase transition in the negative pressure
side of its phase diagram.

In this paper, we present a detailed analysis of the electronic
structure modification across the volume collapse, studied from
first-principles, by means of an explicitly correlated many-body wave
function and accurate 0K quantum Monte Carlo (QMC) techniques.
Remarkably, we have been able to stabilize two
distinct coexisting solutions, $\alpha$ and $\gamma$, with
the full set of structural parameters across the transition
seamlessly connected to the experimental values
at finite $T$.
We prove that the transition results from
a subtle competition between local Coulomb repulsion and bandwidth, the latter determined 
mainly by the $a_\textrm{1g}$ and $t_\textrm{1u}$ atomic orbitals.
The key role is played by the $p$-$f$ hybridization, set by the
octahedral crystal field, which allows the $t_\textrm{1u}$ orbital
to breath between the two phases. In the 
$\gamma$ phase, the chemical bond has weaker $a_\textrm{1g}$
and stronger $t_\textrm{1u}$ channels, due
to more extended $t_\textrm{1u}$ orbitals, if compared to the $\alpha$
phase at the same volume.
This weakens the bond strength while it reduces
the on-site Coulomb repulsion, resulting in a stabilization of the
$\gamma$ phase at larger volumes.

In our approach, the two phases are described by a
paramagnetic Jastrow-correlated Slater determinant (JSD) wave function
sampled by QMC techniques:
\begin{equation}
\Psi_\textrm{JSD}({\bf R}_\textrm{el}) = \exp[-J({\bf R}_\textrm{el})]
\det[\phi_i( \vc{r}_j) ], 
\label{JSD}
\end{equation}
where $1 \le i,j \le N$,  ${\bf R}_\textrm{el}=\{ {\bf r}_1\ldots,
{\bf r}_{N} \}$  is the many-body $N$-electron configuration, and the determinant is 
factorized in two spin components $\uparrow$ and $\downarrow$, since 
 the molecular orbitals $\phi_i$ have a definite 
spin projection along $z$.
Both $J$ and $\phi$ are analytic functions
with parameters that minimize the energy of the
scalar-relativistic first-principles Hamiltonian  (see
\cite{additional} for details). The full Coulomb
electron-ion interaction is replaced by a scalar-relativistic
Hartree-Fock energy
consistent pseudopotential\cite{dolg_private} with $5s^2 5p^6 6s^2
5d^1 4f^1$ atomic reference configuration, which includes semi-core states.

\begin{figure}[hb]
\includegraphics[width=\columnwidth]{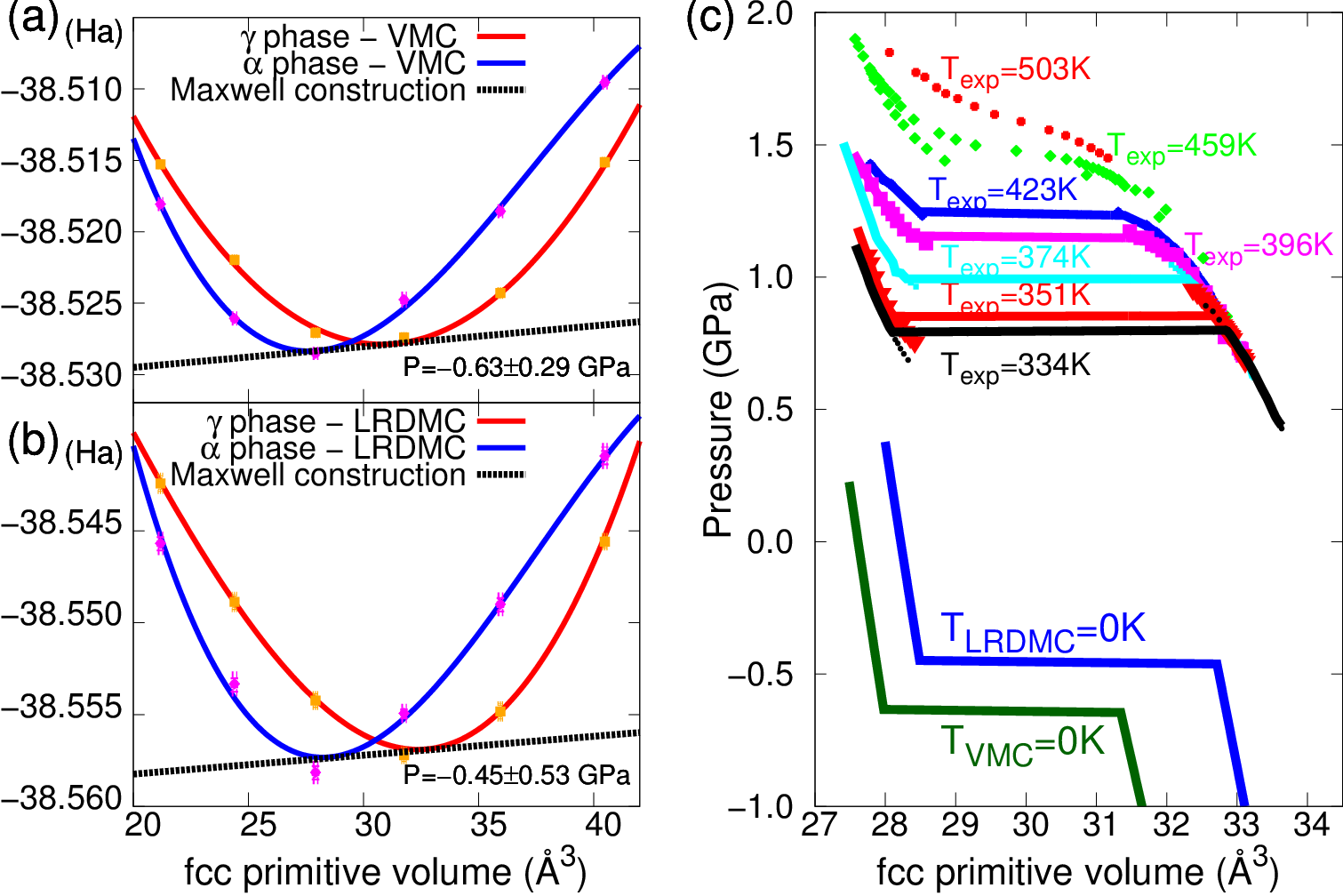}
\caption{
Panels (a) and (b): Equation of states of the $\alpha$ and $\gamma$
phases obtained by VMC and LRDMC calculations, respectively. The black dashed straight line is the
Maxwell construction with the corresponding calculated $\alpha$-$\gamma$
transition pressure. Panel (c): Clapeyron
diagram obtained at 0K in quantum Monte Carlo, compared to the
experimental phase diagram by Decremps et al.\cite{decremps} 
at finite T. Remarkably, the upper and lower critical
volumes are within the experimental range of the coexistence region. 
}
\label{eos}
\end{figure}

The Jastrow factor takes into account strong local
correlations as well as intersite correlations, and thoroughly
modifies the DFT generated Slater determinant. We fully optimized the
JSD wave function in a 32-atom cubic supercell with periodic boundary
conditions, which 
yields structural parameters close to the thermodynamic limit (see
Supplemental Material  in ~\cite{additional}). 
By determining the
variational energy as a function of
the unit cell volume, we evaluated the equation of states at the
variational Monte Carlo level (VMC), as reported in Fig.~\ref{eos}(a). The
\emph{fcc} equilibrium volume per atom $V_\textrm{eq}$ of the $\alpha$ phase
turns out to be 27.4\AA$^3$, in a greatly better 
agreement with the experimental value (28.52\AA$^3$\cite{Koskenmaki1978})
than LDA or GGA DFT calculations. To further improve the electronic structure, we used the
lattice regularized diffusion Monte Carlo (LRDMC) method\cite{lrdmc,lrdmc2}.
In the LRDMC, the starting
point is our best VMC wave function given by (\ref{JSD}), that is projected 
to the ground state with the approximation of 
the fixed nodes, pinned to the ones of the VMC wave function to cope
with the sign problem arising in the imaginary time projection. The
LRDMC equation of states plotted in Fig.~\ref{eos}(b) yields an
equilibrium volume of 28.4\AA$^3$, in 
very good agreement with experiment, while
the bulk modulus $B$ is overestimated
(see Tab.~\ref{phase_transition}).

\begin{table}[ht]
\begin{ruledtabular}
\begin{tabular}{ l l l l}
               & VMC (T=0K) & LRDMC (T=0K)  & exp\\
\hline

$V^\alpha_\textrm{eq}$ (\AA$^3$)  &  27.4 $\pm$ 0.1 &  28.4 $\pm$ 0.2 &
28.52\cite{Koskenmaki1978}\\
$V^\gamma_\textrm{eq}$ (\AA$^3$) & 30.8 $\pm$ 0.2 & 32.3 $\pm$ 0.3 & 34.35\cite{Beaudry1974225} \\
B$^\alpha$ (GPa) & 48 $\pm$ 1 & 50 $\pm$ 3  & 35\cite{Staun1993} \\
B$^\gamma$ (GPa) & 38 $\pm$ 1 & 45 $\pm$ 3 & 21-24\cite{GschneidnerJr1964275,Olsen1985129} \\
\hline
$V_\textrm{min}$ (\AA$^3$) & 28.0 $\pm$ 0.2 & 28.5 $\pm$ 0.3 & 28.2\cite{decremps} \\
$V_\textrm{max}$ (\AA$^3$) & 31.3 $\pm$ 0.3 & 32.7 $\pm$ 0.4 & 32.8\cite{decremps}\\
$\delta V (\%)$ & 11.7 $\pm$ 0.6  &  13.8 $\pm$ 1.1  & 15.1\cite{decremps} \\
$p_t$ (GPa) &  -0.63 $\pm$ 0.29 &  -0.45 $\pm$ 0.53 & 0.7\cite{decremps}\\
$\Delta U$ (meV) & 13 $\pm$ 1  & 12 $\pm$ 3  & 25\cite{decremps}\\ 
\end{tabular}
\caption{
Structural and phase transition parameters for $\alpha$ and $\gamma$ phases
  obtained by VMC and LRDMC, compared with
  the experiment. The $\alpha$-$\gamma$ phase transition parameters
  are taken from Ref.~\onlinecite{decremps} at $T=334$K.
A further detailed comparison with alternative ab-initio methods is reported in
the Supplemental Material.\cite{additional}
}
\label{phase_transition}
\end{ruledtabular}
\end{table}

By starting from the optimal JSD wave function for the $\alpha$ phase, we performed
VMC energy minimizations at much larger volumes ($>$ 40\AA), where we
stabilized a second paramagnetic solution, lower in energy than the $\alpha$
phase. This second solution holds out even at smaller volumes, although
at higher energies. Our computer simulations then
reproduced what is seen in experiment, with a clear hysteresis between
the two states as a function of volume (see Fig.~\ref{eos}(c)). Further
analysis, based on the Maxwell common tangent construction and on the geometry
parameters, confirmed that this solution is fully compatible with
the sought $\gamma$ phase. Its LRDMC equilibrium volume is 32.3\AA$^3$
(Tab.~\ref{phase_transition}),
representing a volume collapse of about 13$\%$; its bulk modulus is
softened with respect to the $\alpha$ phase (as seen in experiments); the
(negative) transition pressure $p_t$ is compatible with the extrapolated
transition line to the negative side of the experimental p-T diagram;
and the lower $V_\textrm{min}$ and upper $V_\textrm{max}$
critical volumes are within the 
experimental range of phase coexistence (Fig.~\ref{eos}(c) and
Tab.~\ref{phase_transition}).  

Once the macroscopic parameters are determined,
our theoretical approach is qualified to provide the
microscopic physical origin of the volume collapse transition. 
By QMC methods it is actually possible to directly access
spin and charge fluctuations, through the measure of the spin-spin
and charge-charge 
correlation functions. Here, we define the charge and spin operators on a cerium site
as $\hat{n}_i  = \int_{V(\vc{R}_i)} \! d\vc{r} ~
(\psi_\uparrow^\dagger(\vc{r}) \psi_\uparrow(\vc{r}) +
\psi_\downarrow^\dagger(\vc{r}) \psi_\downarrow(\vc{r})) $ and  $\hat{\sigma}_i  =   1/2 \int_{V(\vc{R}_i)} \! d\vc{r} ~
(\psi_\uparrow^\dagger(\vc{r}) \psi_\uparrow(\vc{r}) -
\psi_\downarrow^\dagger(\vc{r}) \psi_\downarrow(\vc{r}))$,
where  the fermionic field $\psi_\sigma^\dagger(\vc{r})$ ($\psi_\sigma(\vc{r})$) creates
(annihilates) an electron of spin $\sigma$ at the position $\vc{r}$,
and the integral  
is done over a sphere of
radius $R = 2.5$ a.u. around the nucleus $\vc{R}_i$. 
At the volume V=31.73 \AA$^3$, which falls into the experimental
coexistence region, this integration radius gives
$\langle\hat{n}_i\rangle \approx$ 9 electrons per site in both phases, mainly coming
from the $5s^2 5p^6$ semi-core and the $4f$ states, which are the most localized among the
valence electrons described by our pseudopotential. The on-site charge
fluctuations $\langle\hat{n}_i\hat{n}_i\rangle - \langle
\hat{n}_i\rangle \langle \hat{n}_i\rangle$ computed by VMC are 1.32(2) and 1.35(1) for the
$\gamma$ and $\alpha$ phase, respectively. There is no sizable
difference between the two phases. LRDMC does not change this picture.  Moreover, the Jastrow
parameters which control the charge-charge correlations do not change significantly
between the two phases, in accordance with 
the charge-charge correlation function results.
Therefore, no suppression of double occupancies occurs in the
$\gamma$ phase, signaling that the Mott scenario of the $\alpha$-$\gamma$ transition
should be definitely discarded. This is an important conclusion,
considering that the Mott transition has been proposed as a valid
interpretation of the volume 
collapse until very recently\cite{PhysRevLett.74.2335,abrikosov,PhysRevLett.109.146402}.

In the spin sector $\langle\hat{\sigma}_i\rangle =0$, because the
$\alpha$ and $\gamma$ wave functions do not break the 
spin symmetry, as both states are paramagnetic by construction. 
From the experimental point of view, the cerium \emph{fcc} lattice undergoes
the volume collapse between two paramagnetic states at finite
temperature. However, the $\alpha$ and $\gamma$ phases feature a very
different magnetic susceptibility, the former being Pauli-like, the
latter of Curie-Weiss type. Early calculations based on the Kondo model
\cite{PhysRevLett.49.1106} and later LDA+DMFT 
studies\cite{PhysRevLett.87.276403,PhysRevLett.87.276404,PhysRevLett.94.036401,PhysRevLett.96.066402} 
explained this difference in terms of Kondo local moment
formations in the $\gamma$ phase, while the effective Kondo
temperature of the DMFT impurity problem is much larger in the
$\alpha$ phase, leaving it in the fully screened singlet state.
Thus, the spin response characterizes the two paramagnetic phases at finite
temperature. At 0K the on-site spin-spin correlation
functions $\langle \hat{\sigma}_i \hat{\sigma}_i\rangle - \langle
\hat{\sigma}_i\rangle \langle \hat{\sigma}_i\rangle$
computed by VMC yield 0.5614(3) and 0.5861(4) for the
$\gamma$ and $\alpha$ phase, respectively. Moreover, the spin-spin
correlations have a very short-range. As in the charge sector,
there is no significant difference between the two phases.
This result 
can be understood
in terms of Kondo physics. T$=0$K is lower than any finite Kondo temperature,
so that both phases are in the fully screened regime\cite{demedici2005}. 

\begin{figure}[ht]
\includegraphics[width=\columnwidth]{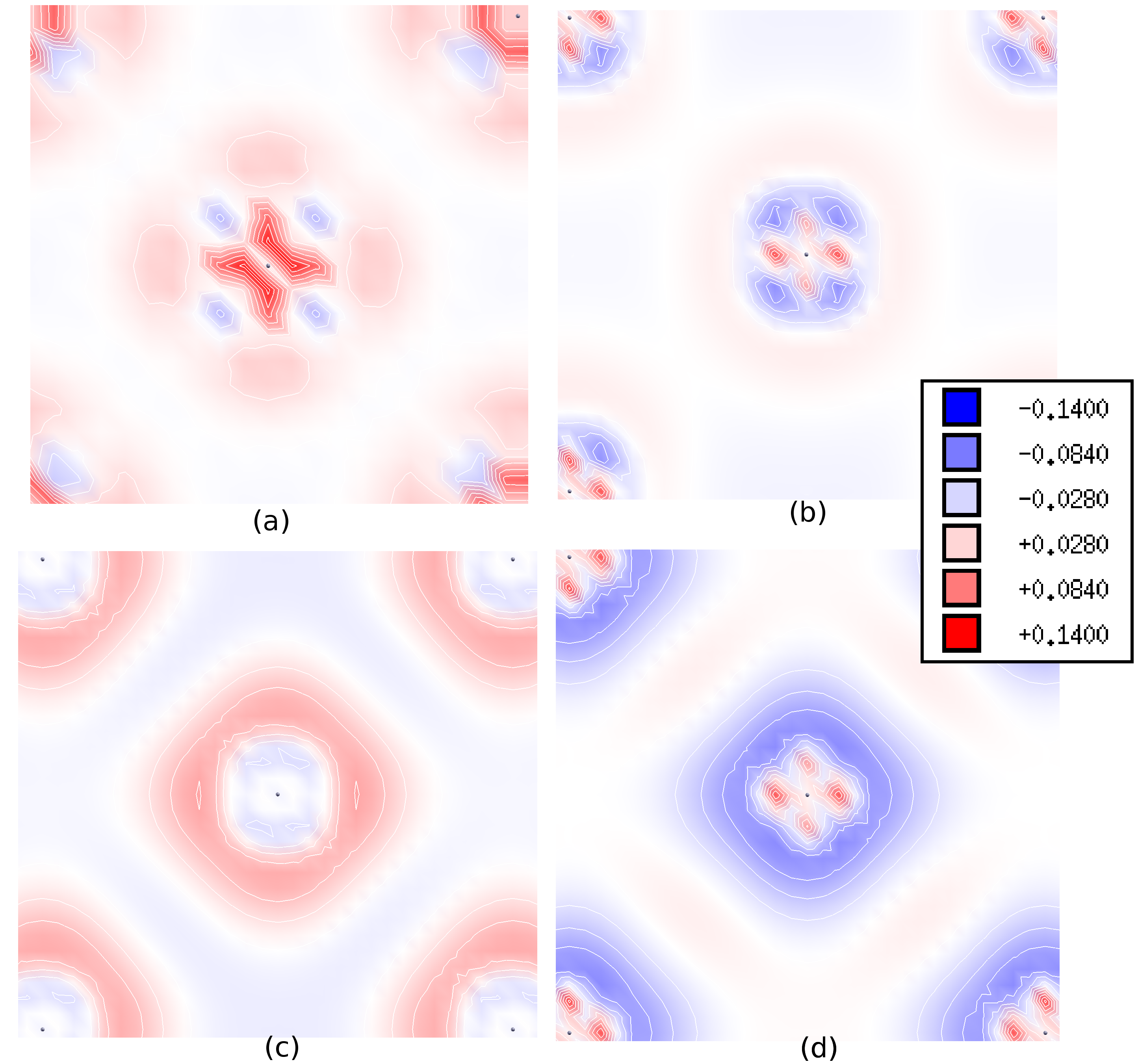}
\caption{
$xy$ contour plot at
$z=0$ (plane containing the central atom).
Panel (a): JSD-VMC density difference $\rho_\alpha - \rho_\gamma$;
Panel (b): $\rho^\textrm{SD}_\alpha - \rho^\textrm{SD}_\gamma$ 
density difference coming from the determinantal part only of the
JSD-VMC wave function. Panels (c), (d): 
$\delta\rho^\textrm{SD}_i=\sum_j \left \{ \rho^\textrm{SD}_\alpha- \rho^\textrm{SD}_\gamma
\right \}_{ij}$, with $\left\{ \rho^\textrm{SD} \right\}_{ij}=\langle
P_i \Psi^\textrm{SD} | \hat{\rho} | P_j \Psi^\textrm{SD} \rangle$ 
the projected electron density, where
$i=\{s+d,~p+f(t_{1u})+f(t_{2u})\}$ are the atomic orbital
symmetries for the (c), and (d) panels,
respectively.
The density values are expressed in \AA$^{-3}$. The JSD-VMC  density
values are twice smaller than the color-code scale printed in the key.
The unit cell volume is 31.73\AA$^3$.
The location of Ce atoms are indicated by gray dots. The
nearest neighbors on the plane of the central atom are at the
square corners. 
}
\label{density_diff}
\end{figure}

The most striking difference between the $\alpha$ and $\gamma$ solutions
is in the electron density $\langle \hat{\rho}(x,y,z) \rangle$,
computed always the the same volume V=31.73
\AA$^3$. The $xy$ contour plot of $\rho_\alpha - \rho_\gamma$ is presented in 
Fig.~\ref{density_diff}(a) for $z=0$, i.e. at the plane containing the
central atom and 4 nearest neighbors.
This shows positive (red) and negative (blue) 
lobes of atomic character. 
The Jastrow factor cannot explain this difference on its own, as we have
seen that its variation between the two phases is weak. Instead, the
difference should come from the Slater determinant $|\Psi^\textrm{SD}\rangle =
\det[\phi]$, once it is combined and 
optimized together with the Jastrow factor in (\ref{JSD}). This is indeed
confirmed by $\rho^\textrm{SD}_\alpha -
\rho^\textrm{SD}_\gamma$, where $\rho^\textrm{SD}=\langle \Psi^\textrm{SD} |
\hat{\rho} | \Psi^\textrm{SD} \rangle$ with $J$ dropped (set to 0) and $\Psi^\textrm{SD}$ frozen, plotted in
Fig.~\ref{density_diff}(b). The charge
density difference carried by the Slater determinant shows a 
similar pattern than the full many-body JSD density.
This is a strong signature that the main difference between the $\alpha$ and $\gamma$
phases at 0K comes from a \emph{static} rearrangement of the electronic
structure, driven by the \emph{dynamic} electron correlation, which -
in our approach - is 
 coded in the Jastrow factor. It is worth pointing out here that both phases 
share almost the same radial charge density (up to a $0.5 \%$
difference) around the nuclei. The main variation 
is in its angular distribution,
suggesting that the transition must be understood in
terms of an electron rearrangement
at the atomic level, which will consequently affect the chemical bond in
the solid.

In order to analyze this hypothesis, we consider  the density matrix 
$D^\textrm{proj}({\bf r},{\bf r^\prime}) = \sum_i \psi^\textrm{proj}_i( {\bf r}) \psi_i ( {\bf r^\prime } )$
left projected over a single cerium atom. This is obtained  
by expanding the molecular orbitals $ \psi_i (\bf r)$  
on an atomic basis set and considering in $\psi^\textrm{proj}_i ({\bf r})$ only
the components referring to the chosen atom. 
We then determine the ``best'' atomic orbitals $\phi^{ANO}_i ({\bf r})$  
representing the projected density matrix by $\sum_{i=1}^k
\phi^{ANO}_i ({\bf r}) \psi_i^R({\bf r^\prime})$ in an optimally
reduced space, namely  
in terms of only $k<<N$ atomic natural orbitals (ANOs) centered on the
reference atom and corresponding 
auxiliary molecular orbitals $\psi_i^R ({\bf r^\prime})$ spanning all the cell.
This can be achieved by a standard Schmidt decomposition, through a  
minimization of the Euclidean distance between the truncated and the projected 
density matrix. The resulting
eigenvalues $\lambda_i^2$  are such that $|D^\textrm{proj}|^2 \approx
\sum_{i=1}^k \lambda_i^2$, and they
are related to the ANOs occupation and
their chemical reactivity (see Supplemental Material in
\cite{additional}). 

\begin{figure}[h]
\includegraphics[width=1.0\columnwidth]{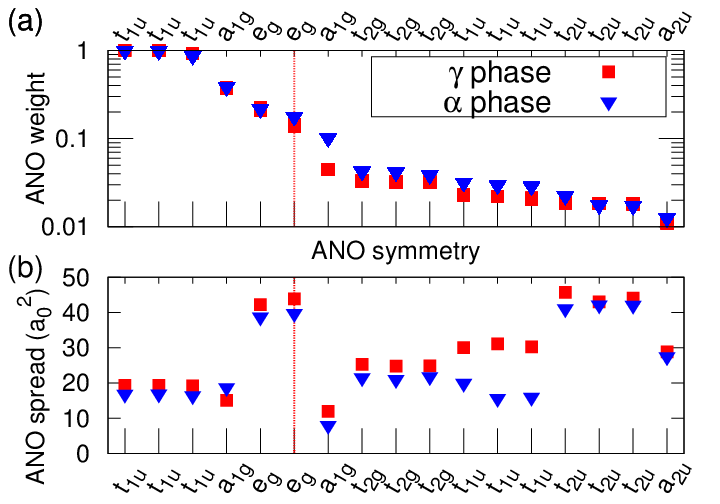}
\caption{
Left panel: First 13 ANO eigenvalues of the density matrix Schmidt decomposition 
for the $\alpha$ and $\gamma$ phases at $V=31.73$ \AA$^3$. Their
symmetry is reported in the x-axis.
The vertical red line represents the filling (6) of a
non-interacting closed-shell pseudoatom in the octahedral field. The variation
between the two phases is
remarkable for the $a_\textrm{1g}$ (atomic $6s$) 7-th eigenvalue,
which is more resonating with the ones below in the 
$\alpha$ phase. 
Right panel: natural orbital localization measured by the spread
$\Omega = \langle \Psi | r^2 | \Psi \rangle - | \langle \Psi | \vc{r}
| \Psi \rangle |^2$. Note that the largest difference comes from the 
$t_\textrm{1u}$ orbitals.  
}
\label{natural_orbitals}
\end{figure}

Any local atomic
variation due to a change in the chemical bond or crystal field is
detected by this approach, as it takes into account the embedding 
of the atom in its environment. In Fig.~\ref{natural_orbitals} we plot
the ANOs $\lambda_i^2$ eigenvalues and their spread, for the same volume as
in Fig.~\ref{density_diff}. 
The first 6 ANOs would be
perfectly occupied in case of non-interacting closed-shell
pseudoatoms. In particular, the first 4 are the
semi-core states.

Two striking features are apparent. 
Firstly, between the $\alpha$ and $\gamma$ ANOs,
there is the variation of the 7-th atomic
orbital weight. In the $\alpha$ phase the 7-th ANO, of
$a_\textrm{1g}$ symmetry,
has almost the same weight as the 5-th and 6-th ANOs, of
$e_\textrm{g}$ symmetry arising from $5d_{3z^2-r^2}$ and
$5d_{x^2-y^2}$ orbitals, degenerate in the octahedral crystal field.
As the isolated atomic ground state
is in the $^1G$ singlet $6s^2 5d^1 4f^1$ configuration (with the $s$ shell full and
inert), it is clear that the s-to-d atomic promotion is crucial to
explain the chemical bond in the $\alpha$ phase, with
the 2 $e_\textrm{g}$ ($5d$) and the $a_\textrm{1g}$ ($6s$) orbitals cooperating 
to set its strength. The cooperative action of $s$ and $d$ orbitals
has been highlighted also in the formation of the Ce dimer\cite{dolg,roos}.

On the other hand, in the $\gamma$ phase, the $a_\textrm{1g}$ ($6s$) ANO
weight is almost an order of magnitude smaller. 
It means that the s-character of the outer-shell region is
weaker in $\gamma$ than in $\alpha$, as shown also by
the density variation 
$\rho^\textrm{SD}_\alpha - \rho^\textrm{SD}_\gamma$ projected onto the
s-atomic orbitals, plotted in Fig.~\ref{density_diff}(c). 

In both $\alpha$ and $\gamma$ phases, above the 7-th $a_\textrm{1g}$ ANO,
there is a series of orbitals 
with small eigenvalues but competing each
other in magnitude. The remaining atomic f-electron shares
a mixed character,
with non-negligible $a_\textrm{2u}$, $t_\textrm{2u}$ and
$t_\textrm{1u}$ occupations.

The second important difference
between $\alpha$ and $\gamma$ is the localization of the
three degenerate $t_\textrm{1u}$ orbitals. The octahedral crystal field makes the
$t_\textrm{1u}$ orbitals strongly hybridized between the $p$ and $f$
atomic symmetries. It turns out that the spread of the $t_\textrm{1u}$
is twice larger in the $\gamma$ phase, as reported in
Fig.~\ref{natural_orbitals}(b). This is consistent with 
Fig.~\ref{density_diff}(d), where the $p+f(t_\textrm{1u})$
projected density is spread over a wider range in the $\gamma$ phase.
The larger $t_\textrm{1u}$ extension reduces the strong local
Coulomb repulsion
and increases the overlap with its neighbors and thus
its bonding character.

To summarize, the sizable difference between the $\alpha$ and $\gamma$
chemical bond character results from
a reduction of the $a_{1g}$ weight together with an increase of
the $t_\textrm{1u}$ overlap. The breathing of the
$t_\textrm{1u}$ orbitals takes place through the hybridization between the
$p$ and $f$ states, coupled by the octahedral crystal field.
The chemical bond in the $\gamma$ phase is weaker (and so the
equilibrium volume is larger) as the $t_\textrm{1u}$-based
bond is less strong than the $s$-$a_{1g}$ one. On the
other hand, the system gains energy by reducing the on-site
Coulomb
repulsion through more extended $t_\textrm{1u}$ orbitals. 

In conclusion, the
volume collapse transition can be understood at 0K   
as a conventional first order transition of electronic origin.
The two phases are well described by the zero-$T$
equation of state, while their relative stability
is provided by tiny entropic effects\cite{decremps}. 
 The underlying mechanism of the volume collapse should
 survive by the addition of the spin-orbit
coupling\cite{PhysRevLett.111.196801}, not present in our
calculations, as in cerium it is much weaker than the local Coulomb
repulsion, although competing to the crystal field splitting\cite{PhysRevB.67.014103}.
Our picture disproves the validity of the Mott model, and puts 
cerium in a quantum phase transition regime. 
Our detailed predictions on the interplay between valence and
localized orbitals can be experimentally tested by
X-ray electron spectroscopy at $L$ edges, to probe the $s$ and
$p$ states.
The electronic phase transition mechanism detailed
in this work can be applied to cerium alloys, and more generally to a wider class of
$f$-electron systems.

We acknowledge useful discussions with S. Biermann, M. Fabrizio, 
Ph. Sainctavit. We are indebted to M. Dolg for providing us with his scalar
relativistic energy consistent tetravalent cerium pseudopotential. 
We thank L. Paulatto for helping us with the generation of cerium PAW
pseudopotentials by using the {\textsc quantum-espresso} atomic ld1 code.
The computational resources used for this work have
been provided by the PRACE grant 2012061116, and IDRIS/GENCI grant
2014096493.  

\bibliography{biblio,cerium}

\end{document}

\title{Supplemental material for the paper: Electronic Origin of the Volume Collapse in Cerium.}

\author{Nicolas Devaux,$^{1}$ Michele Casula,$^{2}$ Fr\'ed\'eric
  Decremps,$^{1}$ and Sandro Sorella$^{3}$}

\address{ $^1$ Institut de Min\'eralogie,
de Physique des Mat\'eriaux et de Cosmochimie,
Universit\'e Pierre et Marie Curie,
case 115, 4 place Jussieu, 75252, Paris cedex 05, France\\
$^2$ CNRS and Institut de Min\'eralogie,
de Physique des Mat\'eriaux et de Cosmochimie,
Universit\'e Pierre et Marie Curie,
case 115, 4 place Jussieu, 75252, Paris cedex 05, France \\
$^3$ International School for Advanced Studies (SISSA) Via Beirut 2,4
  34014 Trieste, Italy and INFM Democritos National Simulation Center,
  Trieste, Italy}

\maketitle

\section{Correlated wave function and QMC methods}

We use the paramagnetic
Jastrow correlated Slater determinant (JSD) as ansatz in all our
calculations, with parameters determined to minimize the energy of the
scalar-relativistic first-principles Hamiltonian. The full Coulomb
electron-ion interaction is replaced by a scalar-relativistic
Hartree-Fock energy
consistent pseudopotential\cite{dolg_private} with $5s^2 5p^6 6s^2
5d^1 4f^1$ atomic reference configuration, which includes the $5s^2 5p^6 $
semi-core states. 

The JSD wave function reads
\begin{equation}
\Psi_\textrm{JSD}({\bf R}_\textrm{el}) = \exp[-J({\bf R}_\textrm{el})]
\det[\psi_i^{MO}( \vc{r}_j^\uparrow)] \det[\psi_i^{MO}( \vc{r}_j^\downarrow)] , 
\label{JSD}
\end{equation}
where $1 \le i,j \le N/2$, and ${\bf R}_\textrm{el}=\{ {\bf r}_1^\uparrow,\ldots,
{\bf r}_{N/2}^\uparrow, {\bf r}_1^\downarrow,\ldots,
{\bf r}_{N/2}^\downarrow\}$  the many-body electron configuration,
with $N$ the total number of electrons.
$\psi_i^{MO}(\textbf{r})$ are molecular orbitals each one  
occupied by opposite spin electrons. 
The orbitals
$\psi_i^{MO}(\textbf{r})$ are expanded in a Gaussian single-particle
basis set $\{\chi^\textrm{det}_j\}$, centered on the atomic nuclei, i.e.
\begin{equation}
\psi_i^{MO}(\textbf{r})=\sum_j \mu_{ij}\chi^\textrm{det}_j (\textbf{r}),
\label{MO_expansion}
\end{equation}
where the sum in the above Equation runs over both the basis set and
nuclear center indices.
The $\{\chi^\textrm{det}_{i}\}$ basis set is uncontracted (primitive) with size of
$7s7p4d5f1g$. The $\mu_{ij}$ and the exponents
of the primitive Gaussian basis set $\{\chi^\textrm{det}_j\}$ are
variational parameters. The first guess for $\psi_i^{MO}$
is provided by density functional theory (DFT) calculations in the
local density approximation (LDA),
performed in the same basis set.

If the HOMO shells are degenerate, as in the case of a 32-atom fcc
supercell with periodic boundary conditions, the symmetry is not
broken thanks to the AGP extension of the JSD ansatz in Eq.(\ref{JSD}). In
the AGP case, the wave function reads
\begin{equation}
\Psi_\textrm{JAGP}({\bf R}_\textrm{el}) = \exp[-J({\bf R}_\textrm{el})]
\det[\phi( \vc{r}_i^\uparrow,\vc{r}_j^\downarrow)]. 
\label{JAGP}
\end{equation}
The function $\phi$ in Eq.(\ref{JAGP}) is written as:
\begin{equation}
\phi({\bf r},{\bf r}^\prime)= \sum_{i=1}^M \lambda_i \psi_i^{MO}({\bf
  r}) \psi_i^{MO}({\bf r}^\prime).
\label{pairing_MO}
\end{equation}
If $M=N/2$, the expansion of Eq.(\ref{pairing_MO}) is equivalent to the
single Slater determinant in Eq.(\ref{JSD}), which factorizes into up
and down components. However, in the case of degenerate shells, $M$
can be larger to include the degenerate
orbitals, with all the HOMO $\lambda_i$ taken equal and tiny. One can prove that
$\Psi_\textrm{JAGP}$ becomes then a linear combination of SDs, each
containing one degenerate orbital. In this way, the shell degeneracy
is correctly taken into account, and the symmetry of the supercell is not broken.

This variational ansatz has been proved very accurate in a large
variety of ab-initio systems,  
molecules\cite{casula_mol,sorella2007,spanu_water} and
solids\cite{sandro_silicon,mariapia_graphene,casula_fese}. 

The Jastrow factor $J$ reads
\begin{equation}
J (\textbf{r}_1, \cdots , \textbf{r}_N) =
\sum_{i=1}^{N_\textrm{nuclei}} \sum_{j=1}^N g^\textrm{1-body}_i({\bf
  R}_i - {\bf r}_j) +  \mathop{\sum_{i,j=1}^N}_{i<j} g( \textbf{r}_i,
\textbf{r}_j ),
\label{J}
\end{equation}
where $g^\textrm{1-body}_i$ is the electron-nucleus term, while $g$
includes electron-electron correlations. 
The one-body part is developed on Gaussian orbitals $\chi_l^i$ (with
$l$ basis set index, and $i$ nuclear index), and it depends on the
i-th nucleus as follows 
\begin{eqnarray}
g^\textrm{1-body}_i({\bf R}_i - {\bf
  r}) & = & (2Z^\textrm{Ce})^{3/4}~u\left((2Z^\textrm{Ce})^{1/4} |{\bf
  R}_i-\vc{r}|\right) \nonumber \\
 & + & \sum_l G_l^i ~ \chi_l^i(\vc{R}_i-\vc{r}), 
\label{Jei}
\end{eqnarray}
where $u(r)=(1+e^{-\alpha
  r})/2\alpha$ is taken to fulfill the 
electron-nucleus cusp conditions with $Z^\textrm{Ce}=12$ the
pseudoatomic number.
The Gaussian basis set is built out of $2s2p2d1f/[1s1p1d1f]$ contracted
$\chi_l^i$ orbitals.
Both $\alpha$, $ G_l^i$, the Gaussian exponents, and the linear coefficients of
the contractions are variational parameters to be optimized. 
The electron-electron part $g$ of the Jastrow factor is defined as
\begin{eqnarray}
g( \textbf{r},
\textbf{r}^\prime)& = & - u^{\sigma,\sigma^\prime}(|\textbf{r}-\textbf{r}^\prime|)
+ \sum_{ijlm} C_{lm}^{ij} ~ \chi_l^i(\vc{R}_i-\vc{r})
\chi_m^j(\vc{R}_j-\vc{r}^\prime) \nonumber \\
 & + & \sum_{ijlm} S_{lm}^{ij} ~ \sigma \sigma^\prime \chi_l^i(\vc{R}_i-\vc{r})
\chi_m^j(\vc{R}_j-\vc{r}^\prime),
\label{Jee}
\end{eqnarray}
where $\sigma$ ($\sigma^\prime$) is the spin $=\pm 1/2$ of the electron at
$\vc{r}$ ($\vc{r}^\prime$), and the homogeneous part
$u^{\sigma,\sigma}(r) = (1-e^{-\beta  r})/4\beta$ 
and $u^{\sigma,-\sigma}(r) = (1-e^{-\beta r})/2\beta$
fulfill the electron-electron cusp conditions for the like- and
unlike-spin particles, respectively. The electron-electron Jastrow
factor in Eq.(\ref{Jee}) contains both charge-charge ($C_{lm}^{ij}$)
and spin-spin ($S_{lm}^{ij}$) correlations.
The Gaussian basis sets
$\chi_l^i$ are the same as the ones in the one-body part.  
$\beta$ and the symmetric matrices $C_{lm}^{ij}$ and $S_{lm}^{ij}$
are variational parameters. The 
basis set parameters (Gaussian exponents and the linear coefficients)
are optimized together with the one-body part (Eq.(\ref{Jei}), and
with the ones of the 
Slater determinant (Eq.(\ref{MO_expansion})). The total number of
parameters in 
$\Psi_\textrm{JSD}$ for the 32-Ce PBC cubic cluster
is about 10000, fully optimized by means of the stochastic
reconfiguration algorithm in Ref.~\onlinecite{srh}.

In order to treat the periodic system in a finite supercell, both the
Jastrow Gaussian basis set $\{\chi_l^i\}$ and the one
$\{\chi^\textrm{det}_j\}$ of the function $\phi$ are made of
periodic Gaussian functions, as described in
Ref.~\onlinecite{sandro_silicon}. Both DFT and quantum Monte Carlo
(QMC) calculations have been carried out using the 
$\textsc{TurboRVB}$ package\cite{turboRVB}.

\section{Finite-size scaling analysis}

All calculations presented in the paper have been done for a supercell
of 32 atoms with periodic boundary conditions (PBC). In order to check
how the PBC 32-atom supercell is close to 
the thermodynamic limit, a finite size scaling analysis has been performed by
extrapolating the QMC data in both $\alpha$ and $\gamma$ phases by
using the method proposed in
Ref.~\onlinecite{PhysRevLett.97.076404_Chiesa}. With respect to other
finite size correction methods, such as the KZK one\cite{kzk}, it has
the advantage of being based on the charge structure factor to correct
for the interaction contributions to the energy. As the structure
factor is an intrinsic property of the many-body system, and it can be
computed at the Monte Carlo level on the variational wave function, it does not
rely on DFT-based external corrections, which can be affected by
the error of the functional. On the other hand, the one-body part of
the kinetic contribution to the total energy is still corrected
at the DFT-LDA level.

The extrapolated values are
reported in Tab.~\ref{phase_transition} for the equilibrium volume per atom
$V_\textrm{eq}$ and the bulk modulus $B$.
The 32-cerium PBC cluster is large enough to get
converged equilibrium volumes within 0.2 \AA$^3$ for the $\alpha$
phase, while slightly larger errors are found in the $\gamma$ phase.
The bulk modulus grows further in the thermodynamic limit, confirming
the tendency for a larger curvature of the QMC equation of
states if it is compared to experiment. To highlight
the accuracy reached by the present results, we recall the reader
that $V^\alpha_\textrm{eq}$ turns out to be 23.56 and 26.80 \AA$^3$ in
LDA and GGA DFT calculations, respectively.

\begin{table}[ht]
\begin{ruledtabular}
\begin{tabular}{ l l l l}
               & VMC (T=0K) & LRDMC (T=0K)  & exp\\
\hline

$V^\alpha_\textrm{eq}$ (\AA$^3$)  &  27.8 $\pm$ 0.1 &  28.2 $\pm$ 0.1 &
28.52\cite{Koskenmaki1978}\\
$V^\gamma_\textrm{eq}$ (\AA$^3$) & 29.5 $\pm$ 0.1 & 30.6 $\pm$ 0.2 & 34.35\cite{Beaudry1974225} \\
B$^\alpha$ (GPa) & 75 $\pm$ 1 & 73 $\pm$ 3 & 35\cite{Staun1993} \\
B$^\gamma$ (GPa) & 50 $\pm$ 1 & 55 $\pm$ 1 & 21-24\cite{GschneidnerJr1964275,Olsen1985129} \\
\hline
$V_\textrm{min}$ (\AA$^3$) & 28.6 $\pm$ 0.2 & 28.9 $\pm$ 0.2  & 28.2\cite{decremps} \\
$V_\textrm{max}$ (\AA$^3$) & 30.9 $\pm$ 0.3 & 31.9 $\pm$ 0.3 & 32.8\cite{decremps}\\
$\delta V (\%)$ & 7.8 $\pm$ 0.4 &  9.9 $\pm$ 0.7 & 15.1 \cite{decremps} \\
$p_t$ (GPa) &  -2.16 $\pm$ 0.41 &  -1.98 $\pm$ 0.55 & 0.7\cite{decremps}\\
$\Delta U$ (meV) & 31 $\pm$ 1 & 37 $\pm$ 2 & 25\cite{decremps}\\ 
\end{tabular}
\caption{
Structural and phase transition parameters for $\alpha$ and $\gamma$ phases
  obtained by VMC and LRDMC after finite size extrapolation, compared to
  the experiment. The $\alpha$-$\gamma$ phase transition parameters
  are taken from Ref.~\onlinecite{decremps} at $T=334$K. 
}
\label{phase_transition}
\end{ruledtabular}
\end{table}

The phase transition parameters obtained by finite size extrapolation confirm the results obtained in the
32-Ce supercell, namely the softening of the bulk modulus in going
from the $\alpha$ to the $\gamma$ phase, the negative transition
pressure, the $V_\textrm{min}$ and $V_\textrm{max}$
within the experimental range of phase coexistence. 
The energy difference $\Delta U$ is larger in the extrapolated
results than in the 32-Ce supercell, as the $\gamma$ curve is slightly pushed up with respect to
the $\alpha$ equation of states (EOS). However, the hierarchy between the $\alpha$
and $\gamma$ EOS is preserved.
The finite size scaling shows that the 32-Ce supercell is large enough to probe the
physics of the $\alpha$-to-$\gamma$ phase transition.

\section{ Calculation of optimal atomic natural orbitals }

In this Section we show how to compute the atomic natural orbitals (ANOs)
$\phi_i^{ANO} (\vc{r}) $ mentioned in the paper.
As explained in the first Section of these Supplemental Materials, the
QMC molecular orbitals $\psi^{MO}_i(\vc{r})$ are optimized in a finite 
localized basis, where each element $\chi^\textrm{det}_i(\vc{r})$ is defined around a given 
atomic position $\vc{R}_i$ (see Eq.(\ref{MO_expansion})).
For the sake of simplicity we consider that all functions described in this Section 
are real, as it is not difficult to generalize
this derivation to the complex case.

The primitive atomic basis $\chi^\textrm{det}_i(\vc{r})$ is not
constrained by any orthogonalization condition, namely the overlap matrix
\begin{equation}
s_{ij} = \langle \chi^\textrm{det}_i | \chi^\textrm{det}_j \rangle
\label{overlap_matrix}
\end{equation}
is an arbitrary strictly positive definite matrix.
Nevertheless the coefficients $\mu_{ij}$ are determined, during the
QMC energy optimization procedure, in a way that the molecular
orbitals remain orthonormal\cite{jcp_mo}: 
\begin{equation}
 \langle \psi^{MO}_i | \psi^{MO}_j \rangle = \delta_{ij},
\label{orthonormal}
\end{equation}
namely,  $\mu s \mu^\dag = I$. 
We define the $N$-electron left-projected density matrix 
on a single cerium atom in a lattice position $\vc{R}$ by:
\begin{equation}
D_\textrm{proj}(\vc{r},\vc{r}^\prime) = \sum\limits_{k=1}^{N/2}
\sum\limits_{i | \vc{R}_i=\vc{R}} \sum_j \mu_{ki} \mu_{kj} 
\chi^\textrm{det}_i(\vc{r}) \chi^\textrm{det}_j(\vc{r}^\prime)
\label{proj_det_mat} 
\end{equation} 
To shorthand the notation, Eq.(\ref{proj_det_mat}) can be also written in terms
of a matrix $\lambda$:
\begin{equation} 
D_\textrm{proj}(\vc{r},\vc{r}^\prime) = \sum\limits_{ij} \lambda_{ij} 
\chi^\textrm{det}_i(\vc{r}) \chi^\textrm{det}_j(\vc{r}^\prime),
\label{proj_det_mat_short}
\end{equation} 
where $\lambda_{ij}=[\mu^\dag \mu]_{ij}$ if the orbital
$\chi^\textrm{det}_i(\vc{r})$ is such that $\vc{R}_i = \vc{R}$,
while $\lambda_{ij}=0$ if $\vc{R}_i \ne \vc{R}$,
whereas the column index $j$ runs all over the atomic basis.

The projected density matrix in Eq.(\ref{proj_det_mat}) carries
information on the intra-atomic electronic structure affected by inter-atomic
interactions between the site $\vc{R}$ and its environment. The
inter-atomic interactions are explicitly kept by the left-partial
projection of the full density matrix. We found this embedding scheme
particularly effective to determine the best ANOs spanning an optimally
truncated Hilbert space.

We represent the projected density matrix of
Eq.(\ref{proj_det_mat}) in a truncated space spanned by
$p$ terms only, as:
\begin{equation}
{\bar D_\textrm{proj}} (\vc{r},\vc{r}^\prime) = \sum \limits_{k=1}^p \phi_k^{ANO}(\vc{r}) \bar \psi_k(\vc{r}^\prime). 
\label{dp}
\end{equation}
In order to find the best ANOs, we minimize the Euclidean distance
$d=|D_\textrm{proj}-{\bar D_\textrm{proj}}|$ between the original and
the truncated density matrices, defined in $ {\cal R}^3 \times {\cal R}^3$
in such a way that:
\begin{equation} 
d^2 = |D_\textrm{proj}|^2 -2 \sum_k \int \!\! d\vc{r} d\vc{r}^\prime D_\textrm{proj}(\vc{r},\vc{r}^\prime) \phi^{ANO}_k (\vc{r}) 
\bar \psi_k( \vc{r}^\prime) + \sum_k \int \!\! d\vc{r} ~ {\bar \psi}^2_k(\vc{r}), 
\label{expd2}
\end{equation}
where $|D_\textrm{proj}|^2 = \int \!\!  d\vc{r} d\vc{r}^\prime
D^2_\textrm{proj}(\vc{r},\vc{r}^\prime)$, and we assumed that  the optimal atomic orbitals 
are orthonormal. This assumption is without loss of generality, as 
- whatever is the solution for the minimum -  we can always 
orthogonalize the corresponding 
 optimal orbitals $\phi_i^{ANO}$ and get a solution   
written in the same form as in Eq.(\ref{dp}). 
We can then take the variation over all possible unconstrained functions 
$\bar \psi (\vc{r})$ and show that $ {\delta d^2 \over \delta \bar \psi_k (\vc{r})} 
=0$ implies:
\begin{equation}
\bar \psi_k(\vc{r}) = \int \!\! d\vc{r}^\prime 
D_\textrm{proj}(\vc{r}^\prime,\vc{r}) \phi^{ANO}_k (\vc{r}^\prime).
\label{unconstrained_psi} 
\end{equation}
Replacing Eq.(\ref{unconstrained_psi}) into (\ref{expd2}) 
yields:
\begin{equation}
d^2= |D_\textrm{proj}|^2 - \sum_k  \int \!\! d\vc{r} d\vc{r}^\prime  \Gamma ( \vc{r},\vc{r}^\prime) \phi_k^{ANO} (\vc{r}) 
\phi_k^{ANO} (\vc{r}^\prime), 
\label{new_d2}
\end{equation}
where the kernel $\Gamma$ is a symmetric function given by:
\begin{equation}
\Gamma ( \vc{r},\vc{r}^\prime) = \int \!\! d\vc{r}'' D_\textrm{proj}(\vc{r},\vc{r}'') D_\textrm{proj}(\vc{r}^\prime, \vc{r}'').
\label{positive} 
\end{equation}
Thus, in order to minimize $d^2$ one needs to maximize the quadratic
form involving $\Gamma$, with the constraint that the orbitals $\phi_k^{ANO} (\vc{r})$ 
are orthonormal.

From Eq.(\ref{positive}) it follows that the kernel function can be expressed in terms of the 
atomic basis $\{\chi^\textrm{det}_i\}$ restricted around a given
cerium at the selected position $\vc{R}_i=\vc{R}$.
By consequence, also the optimal ANOs can be expanded on the same local basis:
\begin{equation}
 \phi^{ANO}_i (\vc{r}) = \sum\limits_{j| \vc{R}_j=\vc{R}} \bar \mu_{ij} \chi^\textrm{det}_j(\vc{r}),  
\label{ano}
\end{equation}
and are actually the eigenvectors of the kernel function $\Gamma$ with maximum 
eigenvalues $\lambda_i^2 \ge 0$. These eigenvalues must be positive 
because Eq.(\ref{positive}) defines a positive definite symmetric kernel. 
In the non orthogonal finite basis $\{\chi^\textrm{det}_j\}$, this turns into  
the generalized eigenvalue Equation:
\begin{equation}
\left[( \lambda s \lambda^{\dag} ) s\right]_{ij} \bar \mu_{kj} = \lambda_i^2 
\bar \mu_{ki},
\label{solver}
\end{equation}
where the matrix $\lambda$ has been defined through Eq.(\ref{proj_det_mat_short}).
Eq.(\ref{solver}) can be immediately solved by standard linear 
algebra packages, by considering that the overlap matrix $s$ is positive definite. 
After diagonalization the eigenvector coefficients satisfy 
the orthogonality requirement  $\bar\mu s \bar \mu^{\dag}=I$, 
that we have previously assumed. Moreover, the truncation error, i.e. the
residual distance, is $d^2 = |D_\textrm{proj}|^2 - \sum_{i=1}^p \lambda_i^2$.

Because the $\{ \phi^{ANO}_i \}$ basis set is optimal in the sense that
we defined in this Section, it has the advantage of not only
being the best compromise between size and
accuracy, but also carrying the physical
information on the most representative atomic states for a site embedded and interacting with
its own environment. 

\bibliography{biblio,cerium}